\documentclass[useAMS,usenatbib]{mn2e}
\usepackage{times}
\usepackage{epsfig}
\usepackage[usenames]{color}

\title[The kinetic power of jets from ADAFs in radio galaxies]
{The kinetic power of jets magnetically accelerated from advection
dominated accretion flows in radio galaxies}
\author[S.-L. Li and X. Cao]{Shuang-Liang
Li\thanks{E-mail:lisl@shao.ac.cn}, Xinwu
Cao\thanks{E-mail:cxw@shao.ac.cn}\\
Key Laboratory for Research in Galaxies and Cosmology, Shanghai
Astronomical Observatory, Chinese Academy of Sciences,\\ 80 Nandan
RD, Shanghai, 200030, China}

\begin{document}

\date{}

\pagerange{\pageref{firstpage}--\pageref{lastpage}} \pubyear{2009}

\maketitle

\label{firstpage}

\begin{abstract}

There is a significant nonlinear correlation between the Eddington
ratio ($L_{\rm bol}/L_{\rm Edd}$) and the Eddington-scaled kinetic
power ($L_{\rm kin}/L_{\rm Edd}$) of jets in low luminosity active
galactic nuclei (AGNs) (Merloni \& Heinz). It is believed that these
low luminosity AGNs contain advection dominated accretion flows
(ADAFs). We adopt the ADAF model developed by Li \& Cao, in which
the global dynamics of ADAFs with magnetically driven outflows is
derived numerically, to explore the relation between bolometric
luminosity and kinetic power of jets. We find that the observed
relation, $L_{\rm kin}/L_{\rm Edd}\propto (L_{\rm bol}/L_{\rm
Edd})^{0.49}$, can be well reproduced by the model calculations with
reasonable parameters for ADAFs with magnetically driven outflows.
Our model calculations is always consistent with the slope of the
correlation independent of the values of the parameters adopted.
Compared with the observations, our results show that over $60\%$ of
the accreted gas at the outer radius escapes from the accretion disc
in a wind before the gas falls into the black holes. The observed
correlation between Eddington-scaled kinetic power and Bondi power
can also be qualitatively reproduced by our model calculations. Our
results show that the mechanical efficiency $\varepsilon$ $(L_{\rm
kin}=\varepsilon \dot{M}_{\rm bondi}c^2)$ varies from
$10^{-2}\sim10^{-3}$, which is roughly consistent with that required
in AGN feedback simulations.

\end{abstract}


\section{Introduction\label{intro}}

It was found that the mass of the black holes in the centre of
galaxies is tightly correlated with galaxy properties
\citep*[e.g.,][]{t2002, b2009}. The co-evolution of massive black
holes and galaxies is usually ascribed to the so-called AGN
feedback \citep{1998A&A...331L...1S}. There are two ways of the
AGN feedback on the ambient gas, i.e., radiative feedback and
mechanical feedback, which correspond to the cases of high and low
Eddington-scaled accretion rate respectively \citep{f2001, c2006}.
The mechanical feedback is in the form of powerful outflows/jets
produced in the most central small region, which can spread to a
large region. The X-ray cavities and bubbles in the galaxies and
clusters are believed to be blown up through the interaction
between the outflows/jets and the intracluster medium (ICM), which
are carefully studied owing to the high resolution observations of
\textit{Chandra} and \textit{XMM-Newton}
\citep*[e.g.,][]{b2004,a2004,a2006,r2006,m2007}. Search the
literature for the data of X-ray cavities, \citet{m2007} complied
a sample of 15 objects containing the data of mechanical power,
the black hole mass and the nuclear luminosity both in the radio
(5 GHZ) band and in the 2-10 kev band. They found a strong
correlation between Eddington-scaled kinetic power and bolometric
luminosity. A similar correlation between Eddington-scaled kinetic
power and Bondi power was also present for this sample \citep*[see
also][]{a2006}.

ADAFs are optically thin, geometrically thick and very hot as most
releasing gravitational energy being stored in the gas instead of
radiating away, which are believed to be present in most of
low-luminosity AGNs \citep*[e.g.,][]{n1994,n1995}.  Due to their
high temperature and positive Bernouli parameter, ADAFs are prone to
be associated by outflows \citep{b1999}, which are consistent with
numerical simulations and observations
\citep*[e.g.,][]{s2001,g1999,y2003}. \citet{m2007} adopted a simple
self-similar model for ADAFs coupled with outflows developed by
\citet{k2004} to explore the correlations between Eddington-scaled
kinetic power, bolometric luminosity and Bondi power. They found
that the correlations can be qualitatively reproduced by the model
calculations \citep*[see][for the details]{m2007}.

The outflows may probably be accelerated from the accretion flow
by the large-scale ordered magnetic fields threading the accretion
flow \citep*[e.g.,][]{b1982,c2002,s2008,liy2009,l2009}.
\citet{l2009} studied the global dynamics of ADAFs with
magnetically driven outflows. In their calculations, the strength
of the large-scale magnetic fields is assumed to scale with the
gas pressure of the accretion flow. The global structure of the
accretion flow together with the mass loss rate in the outflows
are derived simultaneously, and then the kinetic power of the
outflow is available. In this paper, we adopt the model of ADAFs
with magnetically outflows \citep{l2009} to explore the
above-mentioned correlations given by \citet{m2007}.

\section{Model}\label{equations}

In this work, we adopt the model of ADAFs with magnetically driven
outflows/jets surrounding a non-rotating black hole, which is
suggested by \citet{l2009}. In this model, the physical quantities
of the ADAF are integrated in vertical direction, and therefore
the ADAF can be described by one-dimensional hydrodynamical
equations \citep{n1995}. The torque exerted on the accretion flow
due to the magnetically driven outflow. In principle, the
connection between the accretion disc and the outflow can be
explored if the magnetic fields threading the accretion disc is
known and the vertical structure of the disc is provided as the
boundary condition of the outflow \citep*[e.g.,][]{cs94}. For
simplicity, we assume that the pressure of the magnetic fields at
the disc surface is proportional to the total pressure of the
accretion disc in this work. The mass loss rate in the outflow is
evaluated by assuming that the strength of the magnetic fields
threading the disc to be roughly self-similar above the disc as
that adopted in \citet{b1982}, and the effects of the outflow are
then properly included in the one-dimensional hydrodynamical
equations for the ADAF. Although this treatment is rather
simplified, we believe that it should be sufficient good for
modeling the observed statistic correlations in radio galaxies.

We summarize this model briefly in this section. The basic
equations, i.e., the continuity equation, radial momentum equation,
angular momentum equation and the energy equations are given as
follows.

In the presence of magnetically driven outflows, we need to add some
additional terms related with outflows to the dynamical equations
for a classic ADAF \citep*[see][for the details]{c2002b,l2009}. The
continuity equation describing such an ADAF$+$outflow system is
\begin{equation}
\frac{d}{dR}(2{\pi}R\Sigma v_{\rm R})+4\pi R\dot{m}_{\rm w}=0,
\label{mass}
\end{equation}
where all the physical quantities denote their common meanings,
$\dot{m}_{\rm w}$ is the mass loss rate from unit surface area of
accretion flow due to the outflow.

The radial momentum equation is
\begin{equation}
v_{\rm R} \frac{dv_{\rm R}}{dR}-R(\Omega^2-\Omega_{\rm
K}^2)+\frac{1}{\rho}\frac{dP}{dR}-g_{\rm m}=0, \label{radial}
\end{equation}
where $\Omega$ is the angular velocity of the accretion flow at $R$,
the radial magnetic force $g_{\rm m}={B_{\rm r}^{\rm s}B_{\rm
z}}/{2\pi\Sigma}$ ($B_{r}^{\rm s}$ and $B_{z}$ are the radial and
vertical components of the magnetic fields at the disc surface).

The angular momentum equation reads
\begin{equation}
v_{{\rm}R}\frac{d({\Omega}R^2)}{dR}-\frac{1}{{\rho}HR}\frac{d}{dR}(R^{2}H\tau_{r\varphi})+\frac{T_{\rm
m}}{\Sigma}=0, \label{angular}
\end{equation}
where $T_{\rm m}$ is the magnetic torque exerted on the unit surface
area of the disc due to the presence of the outflows \citep{l1994}.
A fraction of the energy and angular momentum of the accretion flow
is carried away by the outflows, which is equivalent to a torque
exerted on the accretion flow. We can calculate the magnetic torque
$T_{\rm m}$ with energy conservation law, provided the power of the
outflow from unit surface area of the accretion disc is known, which
leads to
\begin{equation}
T_{\rm m}={{2}\over {\Omega}}\left[l_{\rm kin}-{\frac
{1}{2}}\dot{m}_{\rm w}\Omega^{2}R^{2}-\dot{m}_{\rm w}
(\varepsilon_{\rm i}+\varepsilon_{\rm e})\right], \label{t_m}
\end{equation}
were $l_{\rm kin}$ and $\dot{m}_{\rm w}$ are the kinetic power and
the mass loss rate of the outflow from unit surface area of the
accretion disc, $\varepsilon_{\rm e}$ and $\varepsilon_{\rm i}$ are
the specific internal energy of electrons and ions respectively. The
main difference between our calculation of $T_{\rm m}$ and that in
\citet{l1994}'s work is that the internal energy of the gas is
properly included in our calculation, while theirs is only for cold
outflows.

The energy equations for ions and electrons are given by
\begin{equation}
{\rho}v_{{\rm}R}(\frac{d\varepsilon_{\rm e}}{dR}-\frac{P_{\rm
e}}{\rho^2}\frac{d\rho}{dR})-{\delta}q^{+}-q_{\rm
ie}+q^{-}+\frac{2\dot{m}_{\rm w}\varepsilon_{\rm e}}{2H}=0,
\label{energy1}\end{equation} and \begin{equation} {\rho}v_{\rm
R}(\frac{d\varepsilon_{\rm i}}{dR}-\frac{P_{\rm
i}}{\rho^2}\frac{d\rho}{dR})-{(1-\delta)}q^{+}+q_{\rm
ie}+\frac{2\dot{m}_{\rm w}\varepsilon_{\rm i}}{2H}=0,
\label{energy2}
\end{equation}
where, the parameter $\delta$ describes the fraction of the
viscously dissipated energy that goes directly into electrons in the
accretion flow, $q^{+}=-{\alpha}PR{d{\Omega}}/{dR}$ is the energy
dissipation rate per unit volume, and the radiative cooling rate
$q^{-}$ consists of synchrotron, bremsstrahlung, and Compton
coolings \citep*[see,][for details]{m2000}. The last terms in Eqs.
(\ref{energy1}) and (\ref{energy2}) represent the internal energy of
the gas in the accretion flow carried away in the outflows. The
effects of bulk kinetic part of the outflows are included in the
angular momentum equation (\ref{angular}), and the structure of the
ADAF is altered, which leads to changes of $q^+$ in Eqs.
(\ref{energy1}) and (\ref{energy2}). The energy transfer rate
$q_{\rm ie}$ from ions to electrons through Coulomb collisions is
\begin{displaymath}
q_{\rm ie}=\frac{3}{2}\frac{m_{\rm e}}{m_{\rm p}}n_{\rm e}n_{\rm
i}\sigma_{\rm T}c\frac{(kT_{\rm i}-kT_{\rm
e})}{K_{2}(1/\theta_{\rm e})K_{2}(1/\theta_{\rm i})}\rm{ln}\Lambda
\end{displaymath}
\begin{equation} \times\left[\frac{2(\theta_{\rm
e}+\theta_{\rm i})^2+1}{(\theta_{\rm e}+\theta_{\rm
i})}K_{1}\left(\frac{\theta_{\rm e}+\theta_{\rm i}}{\theta_{\rm
e}\theta_{\rm i}}\right)+2K_{0}\left(\frac{\theta_{\rm
e}+\theta_{\rm i}}{\theta_{\rm e}\theta_{\rm i}}\right)\right],
\end{equation}
where the Coulomb logarithm $\rm{ln}\Lambda=20$ \citep{s1983}.

For simplicity, the large-scale magnetic field lines are assumed to
thread the accretion disc, and the strength of the magnetic fields
far from the disc surface along the field line to be roughly
self-similar \citep{l1994},
\begin{equation}
B_{\rm p}(R)\sim B_{\rm pd}(R_{\rm d})\left({\frac {R}{R_{\rm
d}}}\right)^{-\zeta}, \label{b_p}
\end{equation}
where $B_{\rm pd}$ is the strength of the poloidal component of the
field at the disc surface, $B_{\rm p}(R)$ is the field strength at
$R$ along the field line threading the disc surface at $R_{\rm d}$.
The self-similar index $\zeta\ge 1$ is required, because the
configuration of any magnetic fields of the disc being able to
accelerate outflows should have an expanding shape above the disc.
In our model calculations, we have not adopted a detailed
configuration of magnetic fields threading the disc, instead, we use
the parameter $\zeta$ to describe how the poloidal field strength
decline along the field line for simplicity as that adopted by
\citet{l1994}. In the calculations of \citet{l1994}, $\zeta=4$ is
adopted .

For a relativistic jet accelerated by the magnetic field of the
disc, the Alfv\'{e}n velocity at the Alfv\'{e}n point is
\citep{c1986}
\begin{equation}
v_{\rm A}=\frac {B_{\rm p}^{\rm A}}{{(4\pi\rho_{\rm A}\gamma_{\rm
j}^{\rm A})}^{1/2}}, \label{v_A}
\end{equation}
where $B_{\rm p}^{\rm A}$ and $\rho_{\rm A}$ are the poloidal field
strength and the density of the outflow/jet at Alfv\'{e}n point, and
$\gamma_{\rm j}^{\rm A}$ is the Lorentz factor of the bulk motion of
the outflows/jets at the Alfv\'{e}n point.

The mass and magnetic flux conservation along the field line
requires
\begin{equation}
{\frac {\dot{m}_{\rm w}}{B_{\rm pd}}}\simeq {\frac {\rho_{\rm
A}v_{\rm A}}{B_{\rm p}^{\rm A}}}. \label{m_b_cons}
\end{equation}
The outflow/jet can be magnetically accelerated over the
Alfv\'{e}n point all along until the modified fast magnetosonic
surface \citep{li1992,cs94,v2003,v2004}, and the bulk velocity of
the outflow/jet at the Alfv\'{e}n point has reached a significant
fraction of its terminal value \citep*[][]{s2008}. In this work,
we have not derived an outflow solution passing smoothly through
the Alfv\'{e}n and slow/fast magnetosonic points as those done in
the previous works \citep*[e.g.,][]{li1992,cs94,v2003,v2004}. The
Lorentz factor of the outflow/jet is
\begin{equation}
\gamma_{\rm j}^{\rm A}\simeq \left[1-\left({\frac {v_{\rm
A}}{c}}\right)^2 \right]^{-{1\over 2}}. \label{gam_j}
\end{equation}

Combining equations (\ref{b_p})--(\ref{gam_j}), the mass loss rate
in the outflow/jet from the unit surface area of the disc is
\begin{equation}
\dot{m}_{\rm w}=\frac{B_{\rm pd}^2}{4{\pi}c}\left[\frac{R_{\rm
d}\Omega(R_{\rm d})}{c}\right]^{\zeta}\frac{{\gamma_{\rm j}^{\rm
A}}^{\zeta}}{({\gamma_{\rm j}^{\rm A}}^2-1)^{(1+\zeta)/2}}.
\label{mw}
\end{equation}

The origin of the ordered magnetic fields threading the disc is
still unclear. It was suggested that the magnetic fields can be
generated through dynamo processes in the disc
\citep*[e.g.,][]{s1973,1998ApJ...501L.189A,1998ApJ...500..703R}, or
the large-scale external magnetic fields are transported inward by
the accretion flow
\citep*[e.g.,][]{1976Ap&SS..42..401B,1994MNRAS.267..235L,2005ApJ...629..960S}.
For simplicity, we assume the magnetic pressure is proportional to
the gas pressure in the accretion flow:
\begin{equation}
P_{\rm m}=\frac{B^2}{8\pi}=\frac{1-\beta}{\beta}P_{\rm gas},
\label{pmag}
\end{equation}
where $\beta$ is the ratio of the gas pressure to the total
pressure, $B$ is the strength of the magnetic fields in the
accretion flow.

The kinetic power of the outflow/jet is extracted from the ADAF
with magnetic fields. The kinetic power of the outflow at
Alfv\'{e}n point can be calculated when its mass loss rate is
derived,
\begin{equation}
L_{\rm kin}=\int^{R_{\rm out}}_{R_{\rm in}} l_{\rm kin} 4\pi R{\rm
d}R, \label{Lkin}
\end{equation}
where $R_{\rm in}$ and $R_{\rm out}$ are the inner radius and
outer radius of the accretion disc, $l_{\rm kin}$ is the kinetic
power of the outflow accelerated from the unit surface area of the
accretion disc. The cooling of the gases during the acceleration
is neglected in Eq. (\ref{Lkin}). The outflow can still be
accelerated beyond the Alfv\'{e}n point in the outflow
\citep*[e.g.,][]{li1992}. The kinetic power of the outflow from
unit surface area of the accretion flow is:
\begin{equation} l_{\rm kin}=(\gamma_{\rm
j}^{\rm A}-1)\dot{m}_{\rm w} c^{2}+\gamma_{\rm j}^{\rm
A}\dot{m}_{\rm w} (\varepsilon_{\rm i}+\varepsilon_{\rm e})+S,
\label{Fkin}
\end{equation}
where the Poynting flux $S$ at the Alfv\'{e}n point in the outflow
is assumed finally to convert to kinetic power of the outflow. For
simplicity, we estimate the Poynting flux $S$ at the Alfv\'{e}n
point roughly with $S \sim {B_{\rm p}^{\rm A}}^2 v_{\rm A}/(4\pi)$
\citep{s2008}. We believe that it should be sufficient good for
modeling the observed statistic correlations in radio galaxies.

The bolometric luminosity of the ADAF can be calculated as
\begin{equation}
L_{\rm bol}=\int^{R_{\rm out}}_{R_{\rm in}} q^{-} 4 \pi RH{\rm
d}R, \label{Lbol}
\end{equation}
where $q^{-}$ is the radiative cooling rate consisting of
synchrotron, bremsstrahlung, and Compton coolings
\citep{n1995,m2000}.

\section{results}\label{results}

Integrating these equations (\ref{mass})-(\ref{energy2}) from the
outer boundary of the flow at $R_{\rm out}$ inwards the black
hole, we can obtain the global structure of the accretion flow
passing the sonic point smoothly to the black hole horizon. In all
our calculations, we adopt the black hole mass $M=8.8\times
10^{8}{\rm M_\odot}$, which is the mean black hole mass of the
samples given in \citet{m2007}. The conventional values of disc
parameters are adopted as: $\alpha=0.1$, $\delta=0.1$, $R_{\rm
out}=5000R_{\rm g}$ and $v_{\rm A}=v_{\rm K}$ \citep*[see][for the
detailed discussion]{l2009}.

\begin{figure}
\includegraphics[width=7.0cm]{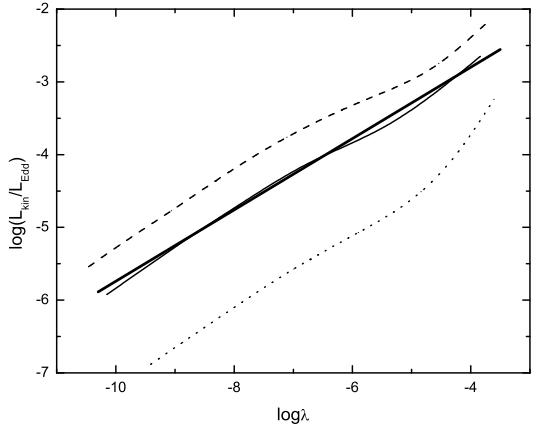}
\caption{The Eddington-scaled kinetic power as functions of
Eddington-scaled bolometric luminosity ($\lambda=L_{\rm bol}/L_{\rm
Edd}$) for different values of $\zeta$. The bold solid line is the
best fitted result of on the correlation givn in \citet{m2007},
i.e., $\textrm{log}(L_{\rm kin}/L_{\rm Edd})=0.49\textrm{log}(L_{\rm
bol}/L_{\rm Edd})-0.78$. The dashed line, solid line and dotted line
are for $\zeta=1$, $\zeta=2$ and $\zeta=4$ respectively.
\label{zeta}}
\end{figure}

\begin{figure}
\includegraphics[width=7.0cm]{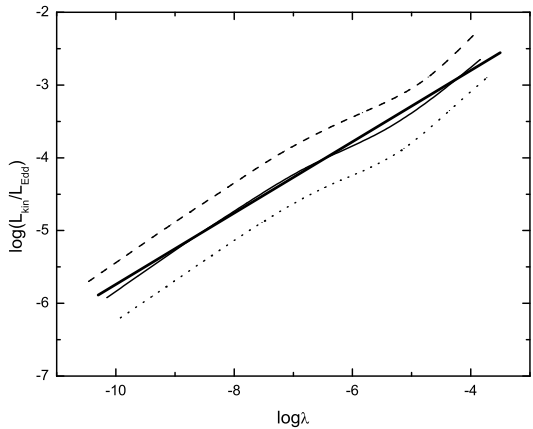}
\caption{The Eddington-scaled kinetic power as functions of
Eddington-scaled bolometric luminosity ($\lambda=L_{\rm
bol}/L_{\rm Edd}$) for different values of $\beta$. The bold solid
line is the best fitted result of on the correlation givn in
\citet{m2007}. The dashed line, solid line and dotted line are for
$\beta=0.9$, $\beta=0.92$ and $\beta=0.94$ respectively.
 \label{beta}}
\end{figure}

\begin{figure}
\includegraphics[width=7.0cm]{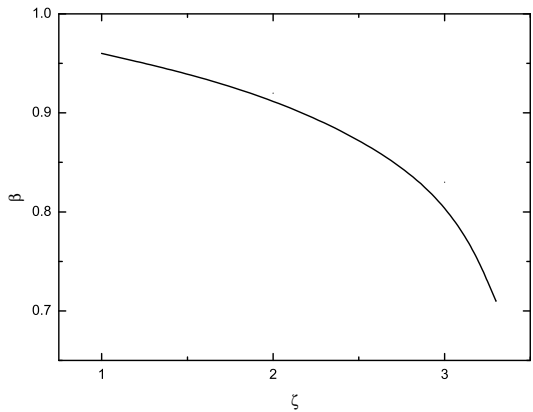}
\caption{The best fitted results to the observed correlation in the
$\zeta$-$\beta$ space.
 \label{beta_zeta}}
\end{figure}

\begin{figure}
\includegraphics[width=7.0cm]{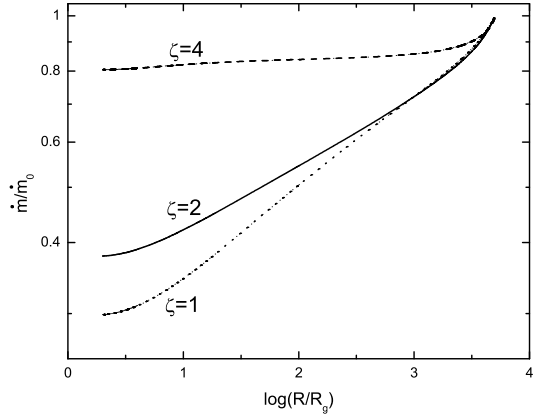}
\caption{The mass accretion rate as functions of radius for
different values of $\zeta$, where $\beta=0.92$ and the accretion
rate at outer radius $\dot{m}_0=10^{-5}$
($\dot{m}_0=\dot{M}_0/\dot{M}_{Edd}$) are adopted. \label{mdot}}
\end{figure}

As described in Sect. \ref{equations}, the structure of the
accretion flow and the outflows is calculated, and then the kinetic
power of outflows and the bolometric luminosity of the accretion
flow are available. Changing the mass accretion rate, the relation
between Eddington-scaled bolometric luminosity $\lambda$
($\lambda\equiv L_{\rm bol}/L_{\rm Edd}$) and kinetic power of the
outflows is plotted in Fig. \ref{zeta} for different values of
$\zeta$ with $\beta=0.92$, i.e., $P_{\rm m}=0.087P_{\rm gas}$, in
which the results are compared with the correlation between $L_{\rm
bol}/L_{\rm Edd}$ and $L_{\rm kin}/L_{\rm Edd}$ given by
\citet{m2007}. The Eddington-scaled kinetic power as functions of
Eddington-scaled bolometric luminosity of the accretion disc is
plotted in Fig. \ref{beta} for different values of $\beta$
($\zeta=2$ is adopted). The observed correlation can be well
reproduced by our model calculations by tuning the values of two
parameters, $\beta$ and $\zeta$. In Fig. \ref{beta_zeta}, the best
fitted model parameters in $\zeta$-$\beta$ space are plotted. We
plot the mass accretion rate as functions of radius for different
values of $\zeta$ in Fig. \ref{mdot}.

\begin{figure}
\includegraphics[width=7.0cm]{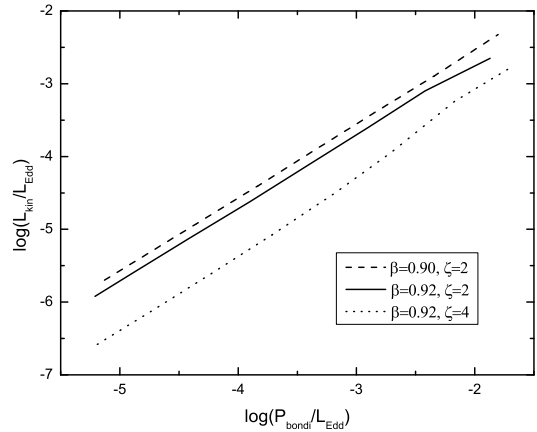}
\caption{The Eddington-scaled kinetic power as a function of
Eddington-scaled Bondi power for different values of $\beta$ and
$\zeta$. \label{bondi}}
\end{figure}

The Bondi power is defined as \citep{m2007}:
\begin{equation}
P_{\rm bondi}=0.1\dot{M}_{\rm bondi}c^{2},\label{Pbondi}
\end{equation}
where $\dot{M}_{\rm bondi}$ is the Bondi accretion rate. The Bondi
accretion rates of the sources in the sample of \citet{m2007} were
estimated with the X-ray observations \citep{a2006}. In this work,
we simply adopt $\dot{M}_0=\dot{M}_{\rm bondi}$, where $\dot{M}_0$
is the accretion rate at the outer radius. We can therefore plot
the dependence of the kinetic powers as functions of Bondi power
in Fig. \ref{bondi} for different values of $\beta$ and $\zeta$.

\section{discussion}\label{conclusions and discussion}

\citet{m2007}'s sample are limited to low luminosity radio galaxies,
which are supposed to be accreting at low rates, probably through a
radiatively inefficient ADAF. In this work, we show that the ADAF
model developed by \citet{l2009} can explain the correlation between
Eddington-scaled kinetic power and bolometric luminosity found by
\citet{m2007}.

In Fig. \ref{zeta}, we show the kinetic power as functions of
bolometric luminosity for different values of $\zeta$, which
describes the distribution of the magnetic field in the space above
the disc. It is found that the slopes of the model calculations with
different values of $\zeta$ are almost constant, which is in good
consistent with the observed correlation (see Fig. \ref{zeta}). The
slope changes very little with different values of $\beta$ (see Fig.
\ref{beta}). This means that the slope of the observed correlation
between $L_{\rm bol}/L_{\rm Edd}$ and $L_{\rm kin}/L_{\rm Edd}$ is
always consistent with our model calculations independent of the
values of parameters adopted. Figure \ref{beta_zeta} shows that a
relatively high $\zeta$ is required for low-$\beta$, which means
that field lines diverse rapidly if the field strength is relative
high, in order to explain the observed correlation between $\lambda$
and $L_{\rm kin}/L_{\rm Edd}$. This provides useful clues to
constructing detailed accretion disc/outflow models, which is beyond
the scope of this work.

We find that the internal energy is not important compared with
other terms in Eq. (\ref{Fkin}). The jet power is mainly related to
the strength of the magnetic fields threading the disc and then to
the disc properties. The bolometric power of the accretion flow is
also related with the disc properties (e.g., density and
temperature). Thus, the observed correlation between the kinetic
power of the outflow and the bolometric power of the accretion flow
can be naturally reproduced by our model calculations. The solid
line in Fig. \ref{mdot} corresponds to the best fitted model
calculation on the observed correlation between Eddington-scaled
kinetic power and bolometric luminosity, which implies that over
$60\%$ of the accreted gas at the outer radius escapes from the
accretion disc in a wind before the gas falls into the black holes
in these low luminosity AGNs. We find that best fitted model
parameters $\zeta$ and $\beta$ are somewhat degenerated, i.e., they
cannot be uniquely determined from the comparison with the
observation, which may be caused by the simplified outflow model
adopted in this work. The observed correlation can be used to
constrain the model if the physics of the origin and the
configuration of the magnetic fields is included, which is beyond
the scope of this work.


The results of our ADAF model with outflows can be qualitatively
explained in the frame of the self-similar ADAF model
\citep{1995ApJ...452..710N}. The radiative efficiency of an ADAF
varies with mass accretion rate. \citet{1995ApJ...452..710N}'s
results showed that the bolometric luminosity of an ADAF $L_{\rm
ADAF}\propto \dot{m}^s$, where $s\sim 2$ for an ADAF due to the
effects of energy advection in the accretion flow. Based on their
self-similar solution for an ADAF, $B^2\propto p_{\rm gas}\propto
\dot{m}$ and then $L_{\rm j}\propto \dot{m}$, which leads to $L_{\rm
j}\propto L_{\rm ADAF}^{1/s}\sim L_{\rm ADAF}^{0.5}$.

Instead of ADAFs discussed above, one may wonder whether the
standard thin accretion disc model can explain this correlation
between $\lambda$ and $L_{\rm kin}/L_{\rm Edd}$, though these
sources are accreting at very low rates. The dependence of jet power
on dimensionless mass accretion rate for standard accretion discs
with different scaling laws for magnetic field strength on the disc
properties were explored in some previous works
\citep*[e.g.,][]{ms96,ga97,l99,c2002}. Most of them suggested that
$L_{\rm kin}/L_{\rm Edd}\propto \lambda$ \citep*[e.g., see][for the
details]{l99,c2002}, and no theoretical model calculations can
reproduce the observed correlation: $L_{\rm kin}/L_{\rm Edd}\propto
\lambda^{0.5}$, based on thin accretion disc models
\citep*[e.g.,][]{ms96,ga97}.

Our calculations also show that the kinetic power increases with
Bondi power when we set the mass accretion rate at the outer radius
$\dot{M}_0=\dot{M}_{\rm bondi}$ (see Fig. \ref{bondi}). This is
roughly consistent with the correlation between $P_{\rm
bondi}/L_{\rm Edd}$ and $L_{\rm kin}/L_{\rm Edd}$ found in
\citet{m2007} (see Fig. 1 in their paper). The mechanical efficiency
$\varepsilon$ $(L_{\rm kin}=\varepsilon \dot{M}_{\rm bondi}c^2)$
varies from $10^{-1}$ to $10^{-5}$ in the AGN feedback simulations
according to the complicated environments and outflows/jets
mechanism, etc \citep{g2009,h2009}. In our calculations, the value
of $\varepsilon$ is about $10^{-2}\sim10^{-3}$, which may be useful
for future  AGN feedback simulations.

\section*{acknowledgements}

We thank the anonymous referee for her/his constructive suggestions
and comments. This work is supported by the NSFC (grants 10903021,
10773020, 10821302 and 10833002), the CAS (grant KJCX2-YW-T03), the
Science and Technology Commission of Shanghai Municipality
(10XD1405000), and the National Basic Research Program of China
(grant 2009CB824800). S.-L. Li thanks the support from the Knowledge
Innovation Program of Chinese Academy of Sciences.

\label{lastpage}

\end{document}